\documentclass[twocolumn,showpacs,amsmath,amssymb,prl,floatfix]{revtex4}

\usepackage{graphicx,bm}

\begin{document}

\title{Asymmetric Franck-Condon factors in suspended carbon nanotube quantum dots}

\author{Fabio Cavaliere$^1$, Eros Mariani$^{2,3}$, Renaud Leturcq$^{4,5}$,
  Christoph Stampfer$^{4,6}$ and Maura Sassetti$^{1}$} 
\affiliation{$^1$
  CNR-SPIN and Dipartimento di Fisica, Universit\`a di Genova,
  Via Dodecaneso 33, 16146 Genova, Italy\\
  $^2$ Institut f\"ur Theoretische Physik, Freie Universit\"at Berlin, Arnimallee 14, 14195 Berlin, Germany\\
  $^3$ School of Physics, University of Exeter, Stocker Road, Exeter, EX4 4QL, United Kingdom\\
  $^4$ Laboratory for Solid State Physics, ETH Zurich, 8093 Zurich,
  Switzerland\\
  $^5$ \mbox{IEMN CNRS-UMR 8520, ISEN, Avenue Poincar\'e, BP 60069, 59652
  Villeneuve d'Ascq Cedex, France}\\
  $^6$  JARA-FIT and II. Institute of Physics, RWTH Aachen University, 52074
Aachen, Germany} 
\date{\today}
\begin{abstract}
  Electronic states and vibrons in carbon nanotube quantum dots have
  in general different location and size. As a consequence, the
  conventional Anderson-Holstein model, coupling vibrons to the dot
  total charge only, may no longer be appropriated in general. Here we
  explicitly address the role of the spatial fluctuations of the
  electronic density, yielding {\em space-dependent} Franck-Condon
  factors. We discuss the consequent marked effects on transport which
  are compatible with recent measurements. This picture can be
  relevant for tunneling experiments in generic nano-electromechanical
  systems.
\end{abstract} 

\pacs{73.23.-b; 85.85.+j; 78.32.-k}
\maketitle 

\textit{Introduction} --- Advances in miniaturization paved the way to
the fabrication of nanodevices in which molecular systems become
active elements of circuits~\cite{cleland}. Tunneling of electrons
through molecules leads to the excitation/de-excitation of quantized
vibrational modes (vibrons) which have been experimentally observed in
suspended carbon nanotubes
(CNT)~\cite{leroy,sapmaz,leturcq,huttel}. Their remarkable electronic
and vibronic properties allowed for the observation of
breathing~\cite{leroy} and stretching vibrons~\cite{sapmaz,leturcq} in
recent transport experiments.\\
In general vibrons couple {\em both} to the total dot charge {\em and}
to the spatial fluctuations of the electron density. The latter
received limited attention so far~\cite{izumida,flensberg,goldwires}. In most
cases the Anderson-Holstein (AH) model~\cite{zazunov,shen} has been
employed, in which the vibron couples {\em only} to the total
charge. The AH model yields {\em position-independent} Franck-Condon
(FC) factors~\cite{fc} which strongly affect
transport~\cite{flensbergold,mitra,koch}. The predicted current
suppression at low bias and the intensity of the vibrational sidebands
have been confirmed in a recent experiment on suspended CNT quantum
dots~\cite{leturcq}.\\
In this paper we show that the effects of density fluctuations are
crucial when the size and location of the dot and of the vibron do not
coincide. They are indeed dramatic when the vibron size $L_{\mathrm
  v}$ is {\em smaller} than the dot size $L_{\mathrm d}$: here, in
sharp contrast with the AH model, {\em position-dependent} FC factors
arise, possibly {\em asymmetric} on the dot tunneling barriers. This
has profound consequences on the transport properties of the
system. Only when $L_{\mathrm v}>L_{\mathrm d}$, the total charge
contribution is dominant and an effective AH model may be
justified~\cite{eros}.\\
\begin{figure}[ht]
\begin{center}
\includegraphics[width=8cm,keepaspectratio]{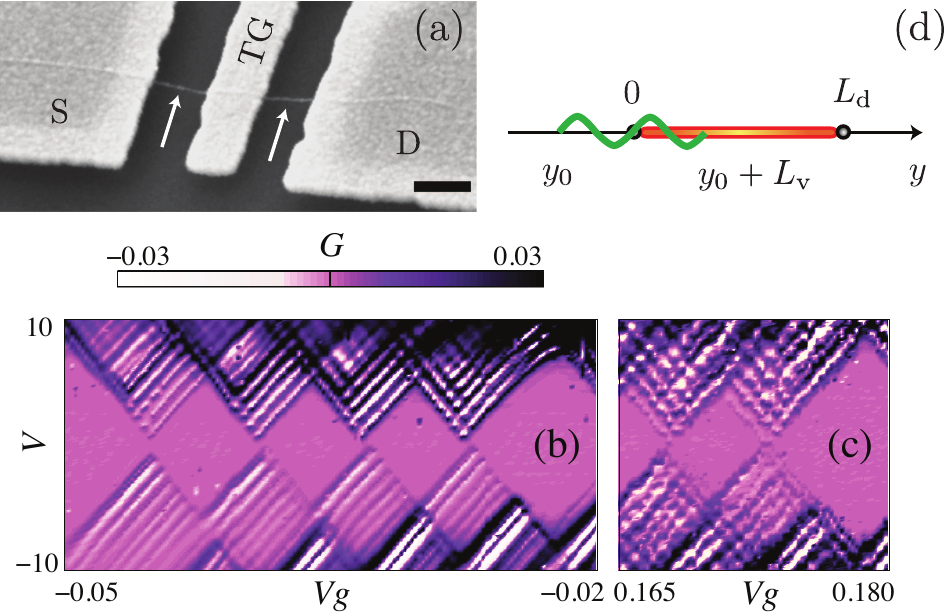}
\caption{(Color online) (a) Scanning electron microscope of the suspended CNT (arrows denote its position) connected to source (S) and
  drain (D) contacts. A top gate (TG) is also present. Scale bar: 200
  nm. (b,c) Experimentally determined differential conductance $G$
  (units $e^{2}/h$) as a function of the top gate voltage $V_{\mathrm
    g}$ (units V) and bias $V$ (units mV). (d) Schematic view of the
  coupled quantum dot-vibron system. The thick part represents the
  quantum dot and the wiggly line the vibron.}
\vskip-0.8cm
\label{fig:fig1}
\end{center}
\end{figure}
Our predictions find an important confirmation in further measurements
on the device considered in Ref.~\onlinecite{leturcq}.  A scanning
electron microscope image, Fig.~\ref{fig:fig1}a, shows the CNT
connected to source (S) and drain (D) leads.  A central suspended
electrode (TG) acts as an electrically insulated top-gate, below which
a quantum dot is formed (for more details see
Ref.~\onlinecite{leturcq}). Transport measurements have been performed
in a pumped $^4$He cryostat with a standard lock-in technique. The
differential conductance $G$ (Figs.~\ref{fig:fig1}b,c) exhibits an
almost perfect fourfold degeneracy in the Coulomb blockade diamonds
and a rich structure of sidebands due to the excitation of stretching
vibrons. The energy of electronic excited states measured on the
Coulomb diamonds yields a dot size $L_{\rm d}\approx 240$ nm, while
the separation of vibrational subbands of about~{0.8 meV} yields
$L_{\rm v}\approx 60\ \mathrm{nm}<L_{\mathrm d}$~\cite{leturcq}.\\
A striking feature is the {\em suppression of vibrational sidebands
  with negative slope} as the gate voltage is varied. While in
Fig.~\ref{fig:fig1}c, with $V_{\rm g}$ in the regime analyzed in
Ref.~\onlinecite{leturcq}, sidebands with both slopes are present, in
Fig.~\ref{fig:fig1}b for $V_{\mathrm g}<0$ those with negative slope
are completely absent.\\
\noindent Here we show that this behaviour requires {\em asymmetric}
FC factors at the tunneling barriers between the dot and the leads. We
stress that such a suppression {\em cannot} be obtained within the AH
model, even assuming strongly asymmetric tunnel barriers. The case of
Fig.~\ref{fig:fig1}c is on the other hand consistent with
quasi-symmetric FC factors, in the spirit of the standard AH model.\\
\noindent In addition, an alternating pattern of positive and negative
differential conductance (PDC/NDC) is observed in all the explored
voltages ranges. This fact will be explained in terms of a
{\em dynamical trapping} of dot states induced by asymmetries in the
tunnel barriers.\\
\indent\textit{CNT Dot-vibron model} --- As a model for our system, we
consider a quantum dot confined between $y_{1}=0$ and
$y_{2}=L_{\mathrm d}$ along the CNT and a vibron clamped at $y_0$ and
$y_0+L_{\rm v}$, with $-L_{\rm v}<y_0<L_{\rm d}$ for a finite overlap
between the two systems (see Fig.~\ref{fig:fig1}d). We describe the
CNT-quantum dot as a Luttinger liquid with two valleys $\eta=\pm 1$
and two spin channels $\sigma=\pm 1$~\cite{egger2} employing standard
bosonization techniques with open boundaries~\cite{fabrizio,yoshioka}
(i.e. the electronic field satisfies $\psi_{\eta,\sigma}
(0)=\psi_{\eta,\sigma}(L_{{\rm d}}^{})=0$). The bosonization picture
is not essential in our analysis, but it simplifies
considerably the formal treatment of the electron-vibron coupling. The dot Hamiltonian is composed
of three terms $H_{\mathrm d}=H_{\mathrm d}^{(0)}+H_{\mathrm
  d}^{(1)}+H_{\mathrm d}^{(2)}$ ($\hbar=1$, $\mu\in\{c+,c-,s+,s-\}$)
\begin{eqnarray}
H_{\mathrm d}^{(0)}&=&\frac{E_{\mathrm c}}{8}(N_{c+}-N_{\mathrm
  g})^2+\frac{\pi v_{\mathrm F}^{}}{8L_{\mathrm
    d}^{}}\left(N_{c-}^{2}+N_{s+}^{2}+N_{s-}^{2}\right)\, ,\nonumber\\
H_{\mathrm d}^{(1)}&=&\frac{1}{2}\sum_{\mu}\sum_{q=1}^{\infty}\left(p_{\mu,q}^{2}+\omega_{\mu,q}^{2}x_{\mu,q}^{2}\right)\ ,\nonumber\\
H_{\mathrm d}^{(2)}&=&\frac{\Delta\varepsilon}{2}(N_{c+}-N_{c-})\, .\nonumber
\end{eqnarray}
The term $H_{\mathrm d}^{(0)}$ describes the energy of $N_{c+}$
electrons in the dot for a given configuration with $N_{\eta\sigma}$
electrons with spin $\sigma$ in branch $\eta$. Here, total (+) and
relative (-) charge ($c$) and spin ($s$) modes have been
introduced~\cite{egger2}, with
$N_{c+}=\sum_{\eta\sigma}N_{\eta\sigma}$,
$N_{c-}=\sum_{\eta\sigma}\eta N_{\eta\sigma}$,
$N_{s+}=\sum_{\eta\sigma}\sigma N_{\eta\sigma}$ and
$N_{s-}=\sum_{\eta\sigma}\eta\sigma N_{\eta\sigma}$. In addition,
$N_{\mathrm g}\propto V_{\mathrm g}$ is the charge induced by the
top-gate voltage $V_{\mathrm g}$, $E_{\mathrm c}$ is the charging
energy and $v_{\mathrm F}$ the Fermi velocity~\cite{prms}. Collective
charge and spin excitations are described as bosonic modes in
$H_{\mathrm d}^{(1)}$. The generalized position and momentum of mode
$\mu$ are respectively $x_{\mu,q}$ and $p_{\mu,q}$, with frequency
$\omega_{\mu,q}=\pi v_{\mu}q/L_{\mathrm d}$ and group velocity
$v_{\mu}$~\cite{prms}.  Finally, $H_{\mathrm d}^{(2)}$ models a shift
between the energy of the two valleys~\cite{Cobden}.\\
\noindent The lowest stretching vibron is described by the harmonic
Hamiltonian $H_{\mathrm{v}}=p_{0}^{2}/2M+M\omega_{0}^{2} x_{0}^{2}/2$,
where $M$ is the vibron mass, $\omega_{0}=\pi v_{\mathrm s}/L_{\rm v}$
its frequency and $v_{\mathrm s}$ the stretching mode
velocity~\cite{prms}. Here, $x_0$ is the amplitude of the lowest
vibron, with distortion field
$u(y)=\sqrt{2}x_{0}\sin{[\pi(y-y_{0})/L_{\rm v}]}$ along the CNT, and $p_0$ is the
conjugate momentum. In a CNT, $v_{\rm s}<v_{\mu}$ and the experimental
estimates yield $\omega_0<\omega_{\mu,1}$~\cite{leturcq,sapmaz}.

\noindent Electrons and vibrations are microscopically coupled via
\begin{equation}
\label{eq:e-ph}
H_{\mathrm{d-v}}=c\!\!\!\!\!\!\int\limits_{\max [0,y_{0}^{}]}^{\min [L_{\mathrm
      d}^{},y_{0}^{}+L_{\mathrm v}^{}]}\!\!\!{\rm d}y\,
[\rho_{\mathrm{R}}^{(c+)}(y)+\rho_{\mathrm{R}}^{(c+)}(-y)]\partial_{y}u(y)\ ,
\end{equation}
where $c$ is the deformation potential coupling
constant~\cite{ando,dress,prms} and we have introduced the total
electron density of right movers
$\rho_{\mathrm{R}}^{(c+)}(y)=\sum_{\eta,\sigma}\psi_{\mathrm{R},\eta,\sigma}^{\dagger}(y)\psi_{\mathrm{R},\eta,\sigma}(y)$
with $\psi_{R,\eta,\sigma}(y)$ their Fermi
operator~\cite{noteob,notedp}.  Notice that, while vibrations couple
to the $c+$ mode only, all four collective electronic modes are
important for transport.  In bosonized form, one has
$\rho_{\mathrm{R}}^{(c+)}(y)=(N_{c+}/2L_{\mathrm
  d})+(1/2\pi)\partial_{y}\phi_{c+}(y)$ with
$\phi_{c+}(y)=\sqrt{\omega_{c+,1}^{}/2}\sum_{q>0}e^{-\xi\pi q/2L_{\mathrm
    d}}[e^{-i\pi qy/L_{\mathrm d}}
  (x_{c+,q}^{}-i\omega_{c+,q}^{-1}p_{c+,q})+\mathrm{h.c.}]$ and $\xi$
the short wavelength cutoff. This expression of the density neglects
the fast oscillating terms due to mixed right and left-moving fermion
fields and is reliable in the large total charge $N_{c+}$ regime with
$N_{c+}\gg L_{\mathrm d}^{}/\pi L_{\mathrm v}^{}$. This condition is
experimentally satisfied in all the ranges of parameters analyzed in
this paper. The coupling Eq.~(\ref{eq:e-ph}) can thus be decomposed
into $H_{\mathrm{d-v}}^{(\mathrm{N})}=c_{0}x_{0}N_{c+}$ and
$H_{\mathrm{d-v}}^{(\mathrm{pl})}=x_{0}\sqrt{M}\sum_{q=1}^{\infty}c_{q}x_{c+,q}$,
due to zero modes and plasmons, respectively. The lengthy but
straightforward expressions of $c_{0}$ and $c_{q}$ will be deferred to
a future publication~\cite{inprep}. We point out that
Eq.~(\ref{eq:e-ph}) accounts for the coupling between vibron and
density fluctuations $H_{\mathrm{d-v}}^{(\mathrm{pl})}$, neglected in
the AH model. The total Hamiltonian $H_{0}^{}=H_{\mathrm d}+H_{\mathrm
  v}^{}+H_{\mathrm{d-v}}^{}$ is thus quadratic in the generalized
coordinates and is diagonalized~\cite{ullersma,mitra} (details will be
given elsewhere~\cite{inprep}) into
\begin{eqnarray}
&&\!\!\!\!\!\!H_{0}^{}=\frac{E_{\mathrm
    c}}{8}(N_{c+}-N_{\mathrm g})^2+\frac{\pi v_{\mathrm
    F}^{}}{8L_{\mathrm
    d}^{}}\left[N_{c-}^{2}+N_{s+}^{2}+N_{s-}^{2}\right]+H_{\mathrm
  d}^{(2)}\nonumber\\
&&\!\!\!\!\!\!+\sum_{\nu\geq
  0}\Omega_{\nu}a_{\nu}^{\dagger}a_{\nu}^{}+\sum_{\mu\neq
  c+}\sum_{\nu\geq
  1}\omega_{\mu,\nu}b_{\mu,\nu}^{\dagger}b_{\mu,\nu}^{}\, .\label{eq:interm}
\end{eqnarray}
The sectors with $\mu\neq c+$ are clearly unaffected by
Eq.~(\ref{eq:e-ph}). On the contrary, in the $c+$ sector new modes,
created by $a_{\nu}^{\dagger}$ with energies $\Omega_{\nu}$
emerge. For $\nu\geq 1$ they represent new collective electron modes
(dressed plasmons), while for $\nu=0$ a vibronic excitation dressed by
plasmons is obtained. The latter is the low-energy vibrational mode
observed in the experiments. The energies $\Omega_{\nu}$ satisfy
$\Omega_{\nu}^{2}=\omega_{0}^{2}+\sum_{q=1}^{\infty}
c_{q}^{2}/\left(\Omega_{\nu}^{2}-\omega_{c+,q}^{2}\right)$, with
$\Omega_{0}^{}<\omega_{0}^{}$ and $\Omega_{\nu}^{}>\omega_{c+,\nu}^{}$
for $\nu\geq 1$ {\em always}. Note that we have reabsorbed a polaron
shift into $E_{\mathrm c}$~\cite{zazunov}.\\ 
\indent\textit{Local FC factors} --- We can now study how the
bosonized Fermi field $\Psi_{R,\eta ,\sigma}(y)$~\cite{yoshioka} is
affected by the transformation above. As we study tunneling at
energies {\em smaller} than the collective charge and spin excitations
of the dot, we restrict the Hilbert space to the $\nu=0$ mode of the
sector $c+$ only. Due to Eq.~(\ref{eq:e-ph}), the vibron operators
$a_{\nu}$ appear in the electronic field, whose truncated form after
the diagonalization reads~\cite{note2}
\begin{equation}
\label{eq:psi}
  \psi_{\mathrm{R},\eta,\sigma}(y)\approx\frac{\chi_{\eta,\sigma}}{\sqrt{2\pi\xi}}e^{-[\lambda_{N}+\lambda_{-}(y)][a_{0}^{\dagger}-a_{0}]}
e^{i\lambda_{+}(y)[a_{0}^{\dagger}+a_{0}]}\, ,
\end{equation}
where $\chi_{\eta,\sigma}$ decreases $N_{\eta,\sigma}$ by one, $\lambda_{N}=c_{0}/\sqrt{2M\Omega_{0}^{3}}$  and 
\begin{equation}
\!\!\lambda_{\pm}(y)=\sqrt{\kappa\frac{\omega_{c+,1}}{\Omega_{0}}}\sum_{q=1}^{\infty}\frac{c_{q}F_{\pm}(y)}{\Omega_{0}^{2}-\omega_{c+,q}^{2}}
\end{equation}
with
$\kappa=1+\sum_{q=1}^{\infty}c_{q}^{2}/(\Omega_{0}^{2}-\omega_{c+,q}^{2})^{2}_{}$
and $F_{\pm}(y)=\sin{\left(\pi qy/L_{\mathrm
    d}+\pi/4\pm\pi/4\right)}$. Note that both $\lambda_{N}$ and
$\lambda_{\pm}(y)$ depend on the CNT and dot parameters and position
only via $y_{0}$, the length ratio $\delta=L_{\mathrm v}/L_{\mathrm
  d}$, the velocities ratio $v_{c+}/v_{s}$, and the dimensionless
coupling $\lambda_{\mathrm{m}}=c/(v_{\mathrm
  s}\sqrt{M\omega_{0}})$~\cite{prms}. The {\em local} FC
factors~\cite{zazunov,mitra} $X^{}_{ll'}(y)=2\pi\xi\left|\langle
N_{\eta,\sigma}-1,l\big|\psi_{{\rm
    R},\eta,\sigma}^{}(y)\big|N_{\eta,\sigma},l^{\prime}_{}\rangle\right|^{2}_{}$
describing tunneling of an electron off the dot while changing the
vibron number from $l$ to $l'$ ($l\leq l'$) have the form
\begin{equation}
\label{eq:fcf}
X^{}_{ll'}(y)=e^{-\lambda^{2}_{}(y)}_{}[\lambda(y)]^{2(l'-l)}_{}
\frac{l!}{l'!}[L_{l}^{l'-l}(\lambda^{2}_{}(y))]^2
\end{equation} 
with
$\lambda^{2}_{}(y)=[\lambda_{N}+\lambda_{-}(y)]^2+\lambda_{+}^{2}(y)$
a {\em position-dependent} effective coupling and $L_{a}^{b}(x)$ the
generalized Laguerre polynomials. This is the main result of our
paper. The position dependence is entirely due to the coupling between
the vibron and the density fluctuations, neglected by the AH model
which instead predicts {\em position-independent} FC factors, with
constant interaction strength $\lambda_{N}$. When $\max
[\lambda_{\pm}(y)]\gg\lambda_{N}$ the position dependence cannot be
neglected, and the AH model becomes questionable. This occurs for
$\delta=L_{\mathrm v}/L_{\mathrm d}<1$ (which is the case of our
experiment) and a vibron located {\em inside} the dot: in this case
indeed $\lambda_{N}=0$.
\begin{figure}[htb]
\begin{center}
\includegraphics[width=8.5cm,keepaspectratio]{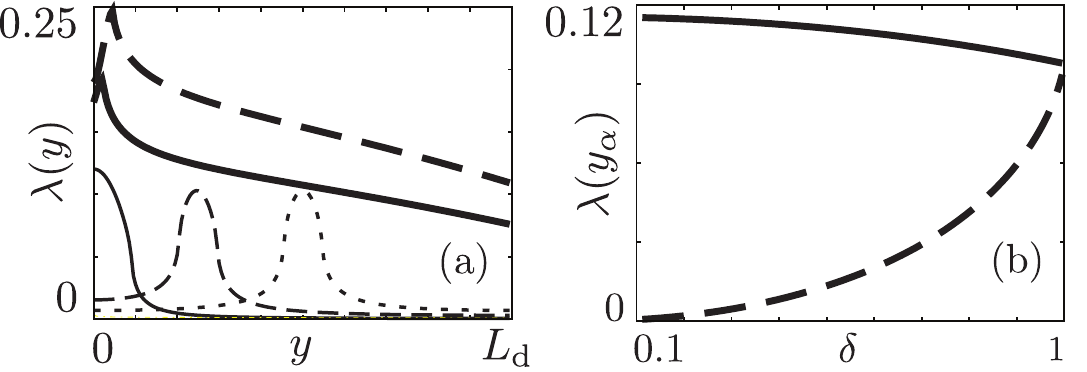}
\caption{(a) Plot of $\lambda (y)$ for $\delta=L_{\rm v}/L_{\rm
    d}=0.1$ and different positions of the vibron center $y_{\mathrm
    c}=y_{0}+L_{\mathrm v}/2$: (thick solid) $y_{\mathrm c}=-L_{\mathrm
    v}/4$; (thick dashed) $y_{\mathrm c}=0$; (thin solid) $y_{\mathrm
    c}=L_{\mathrm v}/2$; (thin dashed) $y_{\mathrm c}=L_{\mathrm d}/4$;
  (thin dotted) $y_{\mathrm c}=L_{\mathrm d}/2$.  (b) Plot of $\lambda
  (y_{\alpha}^{})$ vs. $\delta$ ($\alpha =1,2$) for $y_{0}^{}=0$ and
 $\alpha =1$  (solid) ; $\alpha =2$ (dashed). Notice the strong asymmetry
  for $\delta\ll 1$ and the symmetric $\lambda$'s for $\delta =1$.
  Here, $v_{\mathrm{c+}}/v_{\mathrm{s}}=32$ and
  $\lambda_{\mathrm{m}}=3$ (for a CNT waist $\simeq$ 1
  nm)~\cite{prms}.}
\vskip-0.4cm 
\label{fig:fig2}
\end{center}
\end{figure}
Fig.~\ref{fig:fig2}a shows $\lambda(y)$ for $\delta<1$ and different
locations of the vibron. When the latter sits inside the dot (thin
lines, for $0<y_{0}<L_{\mathrm d}-L_{\mathrm v}$), $\lambda(y)$ is
sizeable only in the vibron region. For vibrons partially outside the
dot (thick lines), $\lambda_{N}\neq 0$ and the position dependence of
$\lambda(y)$ is weaker. For $\delta>1$ (not shown),
$\lambda_{N}\gg\lambda_{\pm}(y)$ which implies
$\lambda(y)\sim\lambda_{N}$, and the {\em spatially-independent} FC
factors of the AH model are obtained~\cite{eros}.\\
\noindent Of particular relevance for transport is the value of the
coupling at the position of the tunneling barriers, $\lambda(y_{1})$
and $\lambda(y_{2})$. For $\delta<1$ and a vibron located
asymmetrically with respect to the dot center, they become very
asymmetric (see the thin solid line of Fig.~\ref{fig:fig2}a), yielding
strongly asymmetric FC factors. In Fig.~\ref{fig:fig2}b,
$\lambda(y_{1,2})$ are shown as a function of $\delta\leq 1$ for a
vibron located near the left barrier. The couplings are strongly
barrier-dependent and vibrational excitations are strongly suppressed
for tunneling on the right. In the symmetric case $\delta=1$, dot and
vibron occupy the same region of space and
$\lambda(y_{1})=\lambda(y_{2})$~\cite{izumida}. Notice however that
$\lambda_{N}=0$.\\
\noindent The maximum value of the coupling for $\delta <1$ is
crucially sensitive to the ratio $v_{\mathrm{s}}/v_{\mathrm c+}$ and
the value of $\lambda_{\mathrm{m}}$.  The coupling of the dot to the
breathing mode reduces $v_{\mathrm c+}$~\cite{egger,izumida},
increasing $v_{\mathrm{s}}/v_{\mathrm c+}$ and allowing to reach
$\lambda(y_{1}^{}) >1$ with $\lambda (y_{2}^{})\ll 1$. In parallel,
recent measurements in graphene~\cite{Bolotin} report a large
deformation potential, which further increases $\lambda_{{\rm
    m}}^{}$.\\ 
\indent\textit{Transport properties} --- In order to address the
electronic transport we introduce the tunneling Hamiltonian coupling
the dot to the leads (represented by the CNT portions outside the dot)
\begin{equation}
H_{\mathrm t}=\sum_{\alpha=1,2}\sum_{\eta,\sigma}
  t_{\alpha,\eta}\psi_{\mathrm{R},\eta,\sigma}^{\dagger}(y_{\alpha})
  \Psi_{\mathrm{R},\eta,\sigma}(y_{\alpha})+{\mathrm{h.c.}}\, ,\nonumber
\end{equation}
where $t_{\alpha,\eta}$ are tunneling amplitudes and
${\Psi}_{\mathrm{R},\eta,\sigma}(y_{\alpha}^{})$ is the right movers
field for lead $\alpha$ . In sequential tunneling, transition rates
are evaluated between eigenstates of $H_{0}$ -
Eq.~(\ref{eq:interm}). For tunneling {\em into} the state $\eta$ of
the dot through the barrier $\alpha$ one has~\cite{koch,haupt}
\begin{equation*}
 \Gamma_{\alpha,\eta}^{\mathrm{(in)}}=\Gamma_{0}
 \frac{|t_{\alpha,\eta}|^{2}}{|t_{\mathrm{2,+1}}|^{2}}
 X_{ll'}^{}(y_{\alpha}^{})f\left[\Delta E+(-1)^{\alpha+1}eV/2\right]
\end{equation*}
where $\Gamma_{0}=2\pi\mathcal{D}|t_{\mathrm{2,+1}}|^{2}/\xi^{2}_{}$
and $\mathcal{D}$ is the leads density of states, while $f(E)$ is the
Fermi function with $\Delta E$ the energy difference between final and
initial dot states. Similar expressions hold for tunnel-out
processes.\\ 
\noindent The experiment allows to estimate the relevant parameters:
$E_{\mathrm c}\approx4.5$ meV (via Coulomb diamonds), the average
$\Gamma_{0}\approx 1\ \mu$eV (via current traces),
$\Omega_{0}\approx\ 800\mu$eV (average vibron sideband separation) and
$k_{\mathrm B}T\approx 90\mu$eV (for $T\approx 1 K$). Since $k_{{\rm
    B}}^{}T\gg\Gamma_{0}^{}$ the sequential regime is justified,
$\Omega_{0}^{}\gg k_{\mathrm B}T$ allows to resolve
vibronic excitations while $\Omega_{0}^{}\gg\Gamma_{0}^{}$ justifies a
rate equation~\cite{koch,noi} neglecting vibronic
coherences~\cite{coherent}. The extremely rich scenario obtained for
different asymmetries of left/right tunnel barriers
$A=|t_{1,\eta}|^2/|t_{2,\eta}|^2$ and of the coupling between leads
and the two valleys $\gamma=|t_{\alpha,-1}|^2/|t_{\alpha,+1}|^2$ will
be discussed in detail elsewhere~\cite{inprep}.
\begin{figure}[ht]
  \begin{center}
\includegraphics[width=8cm,keepaspectratio]{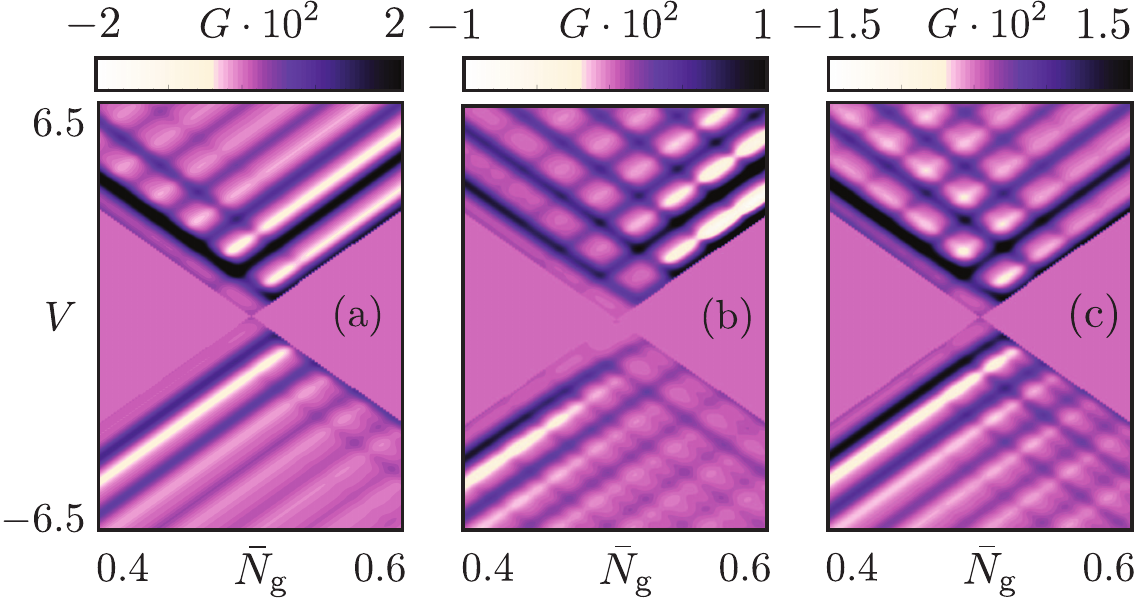}
\caption{(Color online) Plots of the numerical differential conductance $G$ (units
  $e^2/h$) as a function of $\bar{N}_{\mathrm g}=N_{\mathrm g}-3\pi
  v_{\mathrm F}^{}/2E_{\mathrm c}L_{\mathrm d}^{}$ and $V$ (units
  meV). (a) Density plot for $A=1/20$, $\gamma=20$,
  $\lambda^{2}_{}(y_{1}^{})=2.4$, $\lambda^{2}_{}(y_{2}^{})=0.1$; (b)
  same as in (a) but for
  $\lambda^{2}_{}(y_{1}^{})=\lambda^{2}_{}(y_{2}^{})=2.4$; (c) same as
  in (a) but for $A=1/5$, $\lambda^{2}_{}(y_{1}^{})=1.8$ and
  $\lambda^{2}_{}(y_{2}^{})=0.6$. In all panels, $\Omega_{0}=0.8$ meV,
  $k_{\mathrm B}T=0.1\ \Omega_{0}$, $E_{\mathrm c}=4.5$ meV,
  $\Delta\varepsilon=\ 0.48\ \mathrm{meV}$ and
  $\Gamma_{0}=0.8\ \mu\mathrm{eV}$. For simplicity, only one resonance
  is shown.}
\vskip-0.8cm
\label{fig:fig3}
\end{center}
\end{figure}\\
\noindent Here we focus on the relevant case to address the
experimental data in Fig.~\ref{fig:fig1}. For $V_{\mathrm g}<0$ in
Fig.~\ref{fig:fig1}b, we found that the {\em only possible parameter
  range compatible with experimental data} is:
$\lambda(y_{2})\ll\lambda(y_{1})$, $A<1$, $\gamma>1$ and
$\Delta\varepsilon>k_{\mathrm B}T$. The asymmetry of the FC factors is
responsible for the strong suppression of negative-sloped sidebands,
as clearly shown in Fig.~\ref{fig:fig3}a. We want to stress that the
absence of traces with negative slope {\em is not} achievable in the
standard AH model with {\em symmetric} FC factors, even in the
presence of a quite strong asymmetry of the tunneling barriers, as
shown in Fig.~\ref{fig:fig3}b. This proves the need to go beyond the
AH model.\\
\noindent The alternating PDC/NDC traces can be addressed in our model
by the three remaining constraints (on $A$, $\gamma$ and
$\Delta\varepsilon$). The NDC is due to the creation of a bottleneck
in transport: when $\gamma>1$ tunneling into states $\eta=+1$ is
strongly suppressed leading to a dynamical trapping and
NDC~\cite{ciorga,cavaprl}, while the states $\eta=-1$ provides a fast
pathway with ensuing PDC. A shift of the two valleys
$\Delta\varepsilon>k_{\mathrm B}T$ is necessary in order to resolve
the two channels. Finally, the asymmetry $A<1$ allows to obtain the
PDC/NDC pattern in all voltage regimes $V\gtrless 0$.\\
\noindent Analyzing the experimental data, we observe that for
$V_{\mathrm g}>0$ the suppression of conductance traces becomes less
severe (see Fig.~\ref{fig:fig1}c), suggesting more symmetrical FC
factors as in Fig.~\ref{fig:fig3}c, in line with the standard AH
model. In this case NDC traces with both positive and negative slopes
occur for $V>0$, pointing at an asymmetry $A$ weaker than in
Fig.~\ref{fig:fig3}a. The ultimate reason for the relative shift of
electronic vs vibronic wavefunctions at different $V_{g}^{}$ lies in
the unknown details of the electronic and mechanical confinements. Our
predictions could stimulate further developments of experimental
setups with full control over these delicate aspects.

\indent\textit{Conclusions} --- Recent experimental data show the need
of a theory {\em beyond} the usual Anderson-Holstein model of quantum
transport in nano-electromechanical systems. Here we investigate this
new issue by considering the combined role of the electronic charge
and density fluctuations in the coupling to mechanical deformations
for suspended CNT quantum dots.  When vibrons are {\em asymmetrically}
embedded into a {\em larger} dot, {\em position dependent}
Franck-Condon factors arise. The consequent marked effects in the
transport characteristics allow to address experimental features which
could not be captured by the standard AH model.  Our analysis can be
easily extended to consider e.g. planar metallic contacts or the
tunneling from a localized tip.  For small vibrons embedded in larger
dots a spatially-resolved injection of electrons would show a
tunneling suppression sensitive to the vibron location, making our
theory relevant for spatially-resolved scanning tunnelling microscope
measurements as well. Similar effects could be expected also in
systems of higher dimensionality, such as e.g. quantum dots embedded
into suspended graphene sheets.\\
\indent\textit{Acknowledgments} --- F. C. acknowledges support by
INFM-CNR via Seed Project PLASE001.  R. L. and C. S. thank
K. Inderbitzin, L. Durrer, C. Hierold and K. Ensslin for help and
support on the experiment.


\begin{thebibliography}{99}
\bibitem{prms} Typical CNT parameters: $c\approx 30\ {\mathrm{eV}}$,
  $v_{\mathrm{s}}\approx 2.4\cdot 10^{4}\ \mathrm{m/s}$, $v_{\mathrm
  F}=8\cdot 10^{5}\ \mathrm{m/s}$ and $\rho_{0}\approx
  6.7\cdot10^{-7}\ \mathrm{Kg/m^2}$~\cite{dress}. Furthermore,
  $v_{c-}=v_{s+}=v_{s-}=v_{\mathrm F}$ and we assume $v_{c+}\simeq
  v_{{\rm F}}$ due to gate-induced screening.
\bibitem{notedp} Note that Eq.~(\ref{eq:e-ph}) corresponds to
  Eq.~(15) of Ref.~\onlinecite{eros}.
\bibitem{note2} In these expressions we neglected phase factors,
  irrelevant for our analysis.
\end{thebibliography}
\end{document}